\documentclass[aps,pre,twocolumn,superscriptaddress]{revtex4-1}
\usepackage[english]{babel}
\usepackage[utf8x]{inputenc}
\usepackage{amsmath,amsfonts,amssymb,amsbsy,eucal,dsfont}
\usepackage{color}
\usepackage[colorlinks]{hyperref}
\usepackage{graphicx}
\usepackage{rotating}

\newcommand{\hide}[1]{}

\begin{document}

\title{Dynamics of charged gibbsite platelets in the isotropic phase}

\author{Dzina Kleshchanok}
\affiliation{Van 't Hoff Laboratory for Physical and Colloid Chemistry
Debye Research Institute
Utrecht University
H.R. Kruytgebouw, N-701
Padualaan 8, 3584 CH Utrecht,
The Netherlands}

\author{Marco Heinen}
\affiliation{Institute of Complex Systems (ICS-3), Forschungszentrum J\"ulich, D-52425 J\"ulich, Germany}

\author{Gerhard N\"{a}gele}
\affiliation{Institute of Complex Systems (ICS-3), Forschungszentrum J\"ulich, D-52425 J\"ulich, Germany}

\author{Peter Holmqvist}
\affiliation{Institute of Complex Systems (ICS-3), Forschungszentrum J\"ulich, D-52425 J\"ulich, Germany}

\date{\today}

\begin{abstract}
{We report on depolarized and non-depolarized dynamic light scattering, static light scattering, and static
viscosity measurements on interacting charged gibbsite platelets suspended in dimethyl sulfoxide (DMSO). The 
average collective and (long-time) translational self-diffusion
coefficients, and the rotational diffusion coefficient, have been measured as functions of the platelet
volume fraction $\phi$, up to the isotropic-liquid crystal (I/LC) transition.
The non-depolarized intensity autocorrelation function, measured at low scattering wavenumbers,
consists of a fast and a slowly decaying mode which we interpret as the orientationally averaged collective and translational self-diffusion
coefficients, respectively. Both the rotational and the long-time self-diffusion coefficients decrease very strongly,
by more than two orders of magnitude, in going from the very dilute limit to the I/LC transition concentration.
A similarly strong decrease, with increasing $\phi$, is observed for the inverse
zero-strain limiting static shear viscosity. With increasing $\phi$, increasingly strong shear-thinning is observed, accompanied by a
shrinking of the low shear-rate Newtonian plateau. The measured diffusion coefficients are interpreted theoretically in terms of
a simple model of effective charged spheres interacting by a screened Coulomb potential, with hydrodynamic interactions included.
The disk-like particle shape, and the measured particle radius and thickness polydispersities, enter into the model
calculations via the scattering amplitudes. The interaction-induced enhancement of the collective diffusion coefficient by more than a factor of
$20$ at larger $\phi$ is well captured in the effective sphere model, whereas the strong declines both of the experimental translational and 
rotational self-diffusion coefficients are underestimated.}
\end{abstract}

\newcommand{\figureref}{Fig.}
\renewcommand{\figuresname}{Figs.}
\newcommand{\refsname}{References}
\newcommand{\expressionname}{Eq.}
\newcommand{\expressionsname}{Eqs.}

\maketitle

\section{Introduction}\label{sec:Intro}

Colloidal platelets are abundant in nature (\textit{e.g.}, as clay minerals or red blood cells) and can be readily synthesized in the laboratory
in form of mixed metal hydroxides, gibbsite, gold platelets, et cetera. The most prominent examples of colloidal platelets are
various types of natural clays \cite{Clay_Science_Book},
which figure in sediment transport in rivers, and in the oceans and lakes, and which are responsible for dangerous landslides \cite{Khaldoun2009}.
Clays are widely used as rheological modifiers for surface coatings, paints, and drilling fluids \cite{Maitland2000, Murray2000}.

The rheological applications of clays are based on their
microstructural properties, originating from a highly anisotropic shape and a correspondingly anisotropic particle interactions.
As a necessary step towards understanding the more complex behavior of concentrated clay platelet suspensions regarding rheology, sedimentation,
and sediment transport, the fundamental dynamic
properties in dilute isotropic solutions must be addressed.

So far, most studies on the dynamics of clay and colloidal platelet suspensions
have been focused on the non-equilibrium ergodic to non-ergodic transition in dense systems \cite{Ruzicka2004, Nicolai2001, Sciortino2007},
and on the properties of the non-ergodic state \cite{Abou2001, Leheny2004}.
Only few theoretical and experimental investigations have been done so far on less concentrated, 
fluid-state isotropic suspensions of clays or colloidal platelets. For instance,
both the translational and rotational diffusion coefficients of a single platelet were studied 
by simulation, and the resulting data for an extended range of aspect ratios were fitted to general polynomial expressions
for cylinders by Ortega and Garc\'{\i}a de la Torre \cite{DeLaTorre2003}. This study allows 
to compare the single-platelet diffusion properties of non-interacting platelets
with the results of experimental diffusion measurements, but it includes no
hint on how diffusion is affected by particle interactions.

The concentration dependence of sedimentation and diffusion coefficients of uncharged, sterically
interacting platelets and clays has been explored in \cite{Dhont2000,Giannelis2005} by dynamic light scattering and ultracentrifugation.
For platelets, the earlier investigations in \cite{Dhont2000} were made in a rather dilute concentration regime where no effect of particle interactions
on diffusion was detected. Moreover, explicit theoretical expressions have been derived for the 
time-dependent intensity autocorrelation function of non-interacting cylindrical particles such as platelets
\cite{Dhont1996,Dhont2000, Pecora1968, Pecora1990, Kubota1985} which, in principle,
can be used to determine the single-platelet rotational diffusion coefficient from standard dynamic light scattering or
X-ray photon correlation spectroscopy (XPCS) measurements in non-interacting particle systems.

For rather dilute systems of uncharged polymer-grafted
clay suspensions \cite{Giannelis2005}, normalized intensity autocorrelation functions (IACFs) have been measured in 
vertical-vertical (VV) and vertical-horizontal (VH) scattering geometry \cite{BernePecora1976}. A strong slowing of the diffusive modes
was found both in VH and in VV geometry. In VV geometry, only a single diffusive mode was detected, even though it was argued that
a second, cooperative mode should be present due to the osmotic pressure of polymers grafted on the clay particles \cite{Giannelis2005}.

In the present study, we explore how the dynamics of charged colloidal gibbsite platelets suspended in DMSO and present in their isotropic phase,
is affected by direct and hydrodynamic particle interactions. Using static light scattering (SLS) and depolarized and non-depolarized dynamic light scattering (DDLS and DLS),
we study the effect of the platelet concentration
on the measured translational and rotational self-diffusion coefficients, and on the collective diffusion coefficient. Moreover, the concentration- and
shear-rate dependence of the static dispersion viscosity is determined experimentally. Our scattering and rheological experiments cover the full isotropic phase 
concentration regime up to the I/LC transition.

The theoretical description of the platelet dynamics in non-dilute systems is severely complicated by the anisotropy in the direct \cite{Trizac2000}
and, to an even larger extent, by the indirect hydrodynamic interactions (HIs). No readily 
applicable theoretical schemes or simulation results are available for our gibbsite system. Therefore, in a first attempt to analyze theoretically
the experimental trends in the diffusion data, we use a simplifying model where the direct 
interactions in a pair of gibbsite platelets are described approximately
by the spherically symmetric, repulsive part of the Derjaguin-Landau-Verwey-Overbeek (DLVO)
potential \cite{Verwey_Overbeek1948}. Clays, such as Laponite, are mixed silica and metal oxides. This can lead to charges of opposite sign
on their faces and edges \cite{Cohen2001,Harnau2008,Harnau2005}. The present gibbsite platelets consist solely of aluminium hydroxide, so that
their faces and rims of gibbsite have surface charges of the same sign \cite{Franks2006}.
In our effective sphere model of gibbsite in DMSO, we account for the influence of the HIs but leave out the
effects of interaction polydispersity. However, the disk-like shape, and the experimentally determined size polydispersity of the gibbsite platelets,
are accounted for in the scattering amplitudes. The simplifying model of cylindrical platelets with spherically symmetric direct and hydrodynamic interactions
permits us to take advantage of a wealth of theoretical and computer simulation results on rotational \cite{Koenderink2003, Banchio2008} and
translational self-diffusion, and collective diffusion of charged colloidal spheres \cite{Patkowski2005, Banchio2008, heinen2010short}.
The model can be expected to apply approximately to semi-dilute suspensions of isotropically dispersed platelets well below
the overlap volume fraction. For larger concentrations, it is bound to fail.

\section{Experimental details}

We study suspensions of hexagonal colloidal gibbsite [$\gamma$-Al(OH)$_3$] platelets of average radius $\left<R\right> = 44.2$ nm
and average thickness $\left<h\right> = 7.66$ nm, dispersed in dimethyl sulfoxide (DMSO).
The gibbsite particles can be described approximately as a polydisperse system of thin circular cylinders of radius $R$ and (small) height $h$.
We have determined the radius and height distributions from electron microscopy pictures, and using atomic force microscopy,
giving polydispersities (\textit{i.e.}, relative standard deviations) in $R$ and $h$ of $s_R = 17.3\%$ and $s_h = 55.3\%$, respectively.

The mean aspect ratio, $p = \left<h\right> / (2 \left<R\right>) = 0.087$, of platelets is so small that the ultrathin disk limit ($h\to0$) can be 
applied as a reasonable approximation. In this limit, one obtains
$D_t^0 = (D_t^{0,\parallel} + 2 D_t^{0,\perp})/3 = k_B T / (12 \eta_0 \left<R\right>) \approx 3.8\times10^{-12}~\textrm{m}^2/\textrm{s}$ for the
orientationally averaged single-disk translational diffusion coefficient \cite{HappelBrenner1983, KimKarilla1991}, and
$D_r^{0,\perp} = 3 k_B T / (32 \eta_0 {\left<R\right>}^3) \approx 2.2 \times 10^{3} / \textrm{s}$ for the (end-over-end tumbling)
rotational diffusion coefficient determined in small-$q$ DDLS experiments on non-interacting platelets. Here, $k_B$ is the Boltzmann
constant, $T$ is the absolute temperature, and $\eta_0$ is the solvent shear-viscosity.
For $p\to0$, $D_r^{0,\perp}$ becomes equal to the rotational diffusion coefficient, $D_r^{0,\parallel}$, which characterizes rotation
with respect to the platelet rotational symmetry axis. A small but finite value of $p$ lowers somewhat the values of $D_t^0$ and $D_r^{0,\perp}$,
and $D_r^{0,\parallel}$ and $D_r^{0,\perp}$ become different from each other.
Tirado and Garc\'{\i}a de la Torre \cite{Tirado1979, Tirado1980} provide precise polynomial fits 
to their simulation data for the diffusion coefficients of cylindrical platelets as a function of $p$.
However, the aspect ratio range, $p>0.1$, covered by most of these fits, does not include the small aspect
ratio of the present gibbsite platelets. We can alternatively estimate the effect of a non-zero $p$ using analytic results \cite{Perrin1936}
for the single-particle diffusion coefficients of oblate spheroids of same $p$ and same volume $V_p = \pi{\left<R^2 h\right>}$, giving 
$D_r^{0,\perp} = 1.4 \times 10^{3} / \textrm{s}$ and $D_t^0 = 3.1\times10^{-12} \textrm{m}^2/\textrm{s}$.

DMSO is a polar, aprotic, and low-viscous solvent of dielectric constant $\epsilon = 47.2$, at $T = 293 K$, and viscosity
$\eta_0 = 2 \times 10^{-3}$ Pa$\cdot$s, in which the platelets are charge-stabilized, forming a suspension that remains transparent up to the isotropic-liquid crystal (I/LC)
transition. Different from aqueous gibbsite suspensions,
which are turbid already at low concentrations, gibbsite in DMSO systems are transparent and show no multiple scattering in the investigated concentration range.
Moreover, different from aqueous suspensions, there is no residual CO$_2$ contamination, and no solvent self-dissociation, so that
low-ionic strength systems can be easily prepared. 
Thus, we can use DDLS to study the rotational and translational diffusion as a function of the gibbsite number concentration up
to the I/LC transition point. Our standard (D)DLS/SLS apparatus is equipped with a krypton ion laser of wavelength $\lambda_0 = 647$ nm as a light source, and a
$\lambda/2$ plate used as polarizer and analyzer (Bernhard Halle Nachfl., Berlin, Germany). Each sample was measured both in  
vertical-unpolarized (VU) and VH scattering geometry, for values of the wave number, $q$, 
smaller than the value, $q_m$, where the primary peak of the mean scattered intensity, $I(q)$, occurs.

DLS and static viscosity data were recorded for a large number of gibbsite volume fractions $\phi = M/(\varrho_m V)$ up to the I/LC
transition at $\phi_{\text{I/LC}} \approx 8\%$. Here, $M$ is the total mass of added gibbsite of known mass density $\varrho_m$, and $V$
is the suspension volume. In addition, the static shear viscosity, $\eta$, was measured as a function of shear rate $\dot{\gamma}$, for a large number
of concentrations.

\section{Results}

\subsection{Dynamic light scattering results}

\begin{figure*}
\centering
\vspace{-7em}
\includegraphics[width=.6\textwidth,angle=-90]{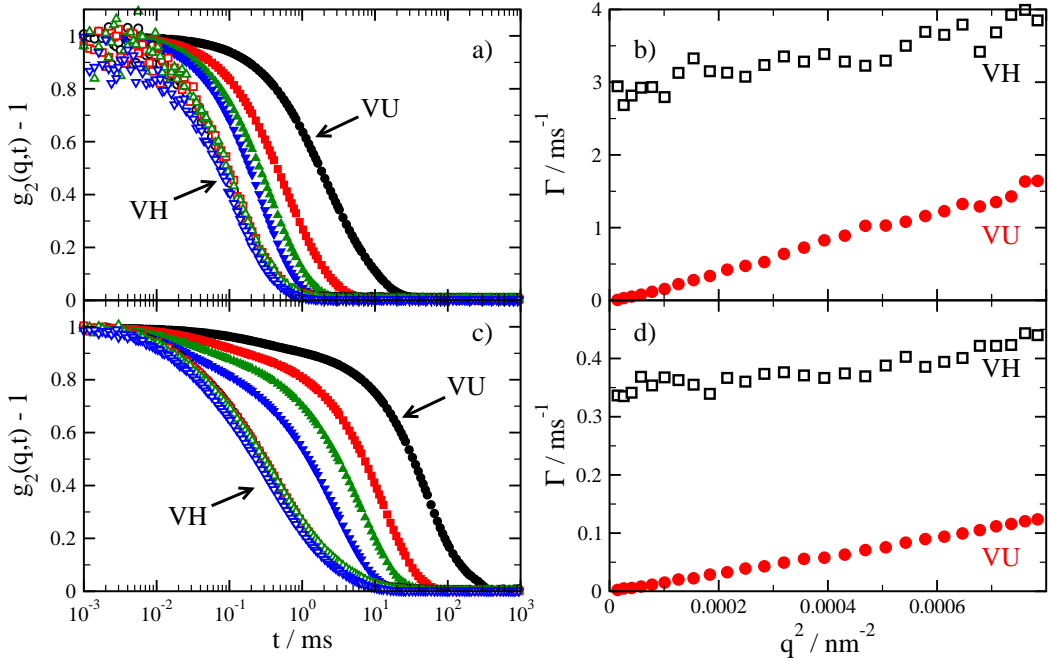}
\caption{Panels a and c: Reduced intensity autocorrelation function, $g_2(q,t)-1$, versus correlation time $t$, for gibbsite in DMSO.
Panels b and d: (mean) relaxation rates, $\Gamma_{VH}$ and $\Gamma_{VU}$, as functions of $q^2$. The gibbsite volume fraction is $\phi = 0.8\%$ in panels
a and b, and $\phi = 4.8\%$ in panels c and d. The wavenumbers in a and c are: $q = 7.6\times10^{-3}$ nm$^{-1}$ (black circles),
$q = 1.47\times10^{-2}$ nm$^{-1}$ (red squares), $q = 2.08\times10^{-2}$ nm$^{-1}$ (green triangles),\
and $q = 2.84\times10^{-2}$ nm$^{-1}$ (blue triangles). Open symbols: VH scattering geometry results.
Filled symbols: $VU \approx VV$ scattering geometry results. 
}
\label{fig:Autocorr_functions}
\end{figure*}

The dynamic light scattering functions in VH and VU geometry have been recorded for a large number of scattering wavenumbers and
concentrations up to the volume fraction at the I/LC transition.
Results for the normalized intensity autocorrelation functions (IACFs), $g_2^{\text{VU}}(q,t)$, (filled symbols) and $g_2^{\text{VH}}(q,t)$ (open symbols),
for two different volume fractions $\phi = 0.8\%$ and $4.8\%$, are shown in \figureref~\ref{fig:Autocorr_functions}. We point out here that
$g_2^{\text{VU}} \approx g_2^{\text{VV}}$ to excellent accuracy, due to the very small optical anisotropy of the gibbsite platelets.
The rather dilute system at $\phi = 0.8\%$ has only weak inter-platelet correlations, whereas the system at $\phi = 4.8\%$ 
is not very far from the I/LC transition.

Consider first the low-concentrated system in panel (a) of \figureref~\ref{fig:Autocorr_functions}, where $g_2^{\text{VU}}(q,t) - 1$
and $g_2^{\text{VH}}(q,t) - 1$ are shown for four different $q$-values located well below $q_m$, with $q\left<R\right> = 1.26$ for
the largest considered $q$. At this low concentration, both IACFs are only slightly stretched, decaying basically
single-exponentially. The stretching is due here to the rather small polydispersity, $s_R = 17.3\%$, in the disk radius $R$.
Due to the small value of $p$, the more pronounced polydispersity in $h$ of value $s_h = 55.3\%$ has only little influence on the particle
diffusion coefficients, and on the scattering amplitudes in the experimentally accessed $q$-range where $q\left<R\right> \lesssim 1$
and thus $q\left<h\right> \lesssim 0.1$. The time relaxation of $g_2^{\text{VH}}(q,t)$ is faster than that of $g_2^{\text{VU}}(q,t)$.

To globally account for polydispersity and particle correlation effects, which both give rise to a potentially 
continuous superposition of correlation times, we have fitted all of the measured functions $g_2(q,t)-1$ in panel (a) of \figureref~\ref{fig:Autocorr_functions}
by the Kohlrausch-Williams-Watts (KWW) stretched exponential form, ${ \{ \exp [ -{(t/\tau_i)}^{\beta_i} ] \} }^2$, characterized by the overall
decay times $\tau_i$ and stretching exponents $\beta_i \leq 1$. An IACF decaying nearly single-exponentially has a value of $\beta$ close to one.
For example, one finds $\beta_i \approx 0.9$ for the IACF's of the low-concentration systems in panel (a) of \figureref~\ref{fig:Autocorr_functions}.         
The KWW mean relaxation rates $\Gamma_i$, with $i \in \{\textrm{VH}, \textrm{VU}\}$, are obtained from
$\Gamma_i = 1/\left[\tau_i\times\Gamma(1/\beta_i)/\beta_i\right]$, where $\Gamma$ denotes the gamma function\cite{Patterson1980}.
In all our experiments, the area coherence factor $C$ in the Siegert relation $C g_1^2(q,t) = g_2(q,t)-1$ is practically equal to one.

In panel (b) of \figureref~\ref{fig:Autocorr_functions}, the resulting relaxation rates $\Gamma_{\text{VH}}$ and
$\Gamma_{\text{VU}}$ at $\phi = 0.8\%$ are plotted as functions of $q^2$.
At the small $q$-values considered where $q\left<R\right> \lesssim 1$, the relaxation rates in VH and VU$\approx$VV geometry can be expressed
as $\Gamma_{\text{VH}} =6 D_r^{\perp} + \mathcal{O}(q^2)$ and $\Gamma_{\text{VV}} = D_t q^2 + \mathcal{O}(q^4)$, respectively
\cite{BernePecora1976, Dhont1996, Pecora1990, Kubota1985, Jones1995}, where $D_t$ and  $D_r^{\perp}$ are interpreted
as the translational and (end-over-end tumbling) rotational diffusion coefficients, respectively.
In this interpretation, polydispersity effects are accounted for in an overall way through the KWW relaxation rates.
Contributions to $\Gamma_{\text{VV}}$ of $\mathcal{O}(q^4)$ which arise from rotational diffusion and rotational-translational
coupling, are of no relevance for the considered small $q$-values.

Regarding the low-concentration system results in panels (a) and (b), no distinction is required between
short- and long-time diffusion coefficients. The slopes and intercepts of $\Gamma_{\text{VU}}$ and $\Gamma_{\text{VH}}$ in panel (b),
obtained from a $q^2\to0$ extrapolation, are approximately equal to the zero-concentration diffusion coefficients.
A zero-$\phi$ extrapolation of all investigated systems, with the relaxation rates obtained as described above, leads to $D_t^0 = 2.1 \times 10^{-12}~\text{m}^2/\text{s}$
and $D_r^{0,\perp} = 0.34\times10^{3} / \textrm{s}$. Both values are somewhat smaller than those obtained from
the ultrathin platelet approximation using $R = \left<R\right>$.

The data for $\Gamma_{\text{VH}}$ in panel (b) show an overall linear increase in $q^2$, but are
more noisy than those for $\Gamma_{\text{VU}}$. This originates from the much lower scattering intensity in VH geometry, which for 
the gibbsite system is by a factor of $10^{-4}$ smaller than that in VU geometry.   
 
At larger platelet concentrations, where direct and hydrodynamic interactions come into play, one notices
interesting changes in the scattering functions.
The normalized VU and VH IACFs, and the corresponding KWW relaxation rates at $\phi = 4.8\%$, are depicted
in panels (c) and (d) of \figureref~\ref{fig:Autocorr_functions}, respectively.
It is apparent from panel (c) that, in VU geometry, an additional quickly relaxing mode occurs.
Moreover, the decay of the slow-mode in $g_2^{\text{VU}}(q,t)-1$ is slightly more stretched. In VH geometry,
no additional mode is seen at this larger concentration, but the decay is more stretched than in the $\phi = 0.8$\% case. 
These features are indicators of pronounced platelet correlations at $\phi = 4.8\%$, requiring now the distinction between short- and long-time
diffusion properties. The subdiffusive transition regime is characterized by the structural relaxation 
time defined by $\tau_I^0 = 1/(6 D_r^{0,\perp}) \approx 0.05$ ms.   

Using again the Siegert relation, we can fit the two-mode decay of $g_2^{\text{VU}}$ observed in panel (c)
in the probed time window by the two-exponential form, 
\begin{eqnarray}\label{eq:double_exponential}
g_2^{\text{VU}}(q,t) - 1 =
{\left[A e^{\displaystyle{-\Gamma_t^{\text{VU}} t}} + (1-A) e^{\displaystyle{-\Gamma_c^{\text{VU}} t}} \right]}^2,
\end{eqnarray}
involving three fit parameters $A, \Gamma_t^{\text{VU}}$ and $\Gamma_c^{\text{VU}}$ with $0 \leq A \leq 1$
and $\Gamma_c^{\text{VU}} > \Gamma_t^{\text{VU}} > 0$.
Both the fast-mode and the slow-mode relaxation rates, $\Gamma_c^{\text{VU}} = q^2 D_c$ and $\Gamma_t^{\text{VU}} = q^2 D_t$, show the expected
diffusive $q^2$-dependence, allowing for the determination of the associated diffusion coefficients $D_c$ and $D_t$, respectively.
For $\phi = 4.8\%$, the ordering relations $D_c > D_t^0$ and $D_t < D_t^0$ are obeyed.
In panel (d), $\Gamma_t^{\text{VU}} \approx \Gamma^{\text{VV}} = q^2 D_t + \mathcal{O}(q^4)$ is plotted as a function of
$q^2$, with $D_t$ inferred from the small-$q$ slope. The corresponding fast mode rate, $\Gamma_c^{\text{VU}}$, is not shown in the figure.
Its associated diffusion coefficient, $D_c$, has been determined as the slope of $\Gamma_c^{\text{VU}}(q^2)$, extrapolated to $q=0$.

Panel (d) shows additionally the relaxation rate, $\Gamma^{\text{VH}} = 6 D_r^{\perp} + D_t q^2$,
of the somewhat stretched out single-exponential decay of $g_2^{\text{VH}}(q,t)$, determined again using the KWW analysis. Note here that the small-$q$ slope
of $\Gamma^{\text{VH}}(q^2)$ is equal, within the experimental noise, to that of $\Gamma_t^{\text{VU}}(q^2)$.
The slope in both geometries is identified as the translational self-diffusion
coefficient, $D_t$, of interacting platelets in the isotropic phase. Likewise, the coefficient $D_r^\perp$ determined from the zero-$q$ intercept
of $\Gamma^{\text{VH}}(q^2)$ in panel (d), obeying $D_r^\perp < D_r^{0,\perp} = D_r^\perp(\phi\to0)$, can be interpreted as the rotational
self-diffusion coefficient at non-zero concentrations.   

\begin{figure}
\centering
\includegraphics[width=.39\textwidth,angle=-90]{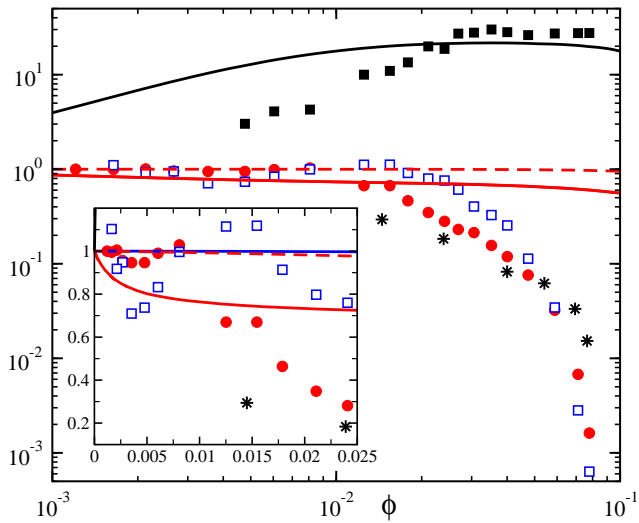}
\caption{Volume fraction dependence of various dynamic properties including the normalized
fast-mode diffusion coefficient, $D_c/D_t^0$, (filled squares), and the slow-mode diffusion
coefficient, $D_t/D_t^0$, (filled circles), both obtained in VU geometry, and the normalized (mean) rotational diffusion
coefficient, $D_r^\perp/D_r^{0,\perp}$  (open squares), obtained from VH measurements.
Stars: Inverse low-shear-rate viscosity $(\eta/\eta_0)^{-1}$, obtained from rheometric measurements. 
Solid black and dashed red lines: Predictions for $D_c/D_t^0$ and $D_t^S/D_t^0$ within the effective charged sphere model,
calculated using the corrected $\delta\gamma$ scheme. Solid blue line (inset only):
Scaling relation $D_r^S/D_r^0 = 1 - 1.3\phi_{\textrm{eff}}^2$ for the short-time rotational diffusion coefficient
of effective charged spheres, using $\phi_{\textrm{eff}} = 1.93\phi$.
Solid red line: $D_t^L/D_t^0$ of effective charged spheres, calculated using the simplified MCT
scheme with far-field HIs correction. The inset magnifies the details at low $\phi$ on a double linear scale.
}
\label{fig:Modes_vs_phi}
\end{figure}

The same evaluation procedure as explicated in \figureref~\ref{fig:Autocorr_functions} for two
specific concentrations, was applied to an extended set of concentrations up to the I/LC transition value.
The obtained reduced diffusion coefficients, $D_t/D_t^0$, $D_r^\perp/D_r^{0,\perp}$ and $D_c/D_t^0$, are depicted in \figureref~\ref{fig:Modes_vs_phi}
in their dependence on $\phi$.
According to \figureref~\ref{fig:Modes_vs_phi}, $D_t$ and $D_r^\perp$ remain constant, within the experimental scatter,
in the lower-concentration range of $\phi \lesssim 0.5\%$. This is the expected behavior of particles at low concentration which diffuse practically uncorrelated.
On the other hand, already at $\phi = 1\%$, the fast-mode coefficient, $D_c$, exceeds largely its zero-concentration value $D_t^0$ whereas, as viewed on
the extended vertical scale of \figureref~\ref{fig:Modes_vs_phi}, $D_t$ and $D_r^\perp$ are just about to start their strong decline 
below their respective infinite dilution values.

A concentration dependence similar to that of $D_c$ has been reported for the collective diffusion coefficient of polymers and flexible
rod systems \cite{Delsanti1977,Ballauf2009}.
However, the gibbsite platelets studied in the present work include no flexible parts. Moreover, the onset of the strong increase of $D_c(\phi)$
is observed at concentrations far below the platelet overlap concentration, $\phi^* = (3/2) p$, of about $13\%$.
Incidentally, an alternative definition of the overlap concentration invoking the random close packing volume fraction, $f = 0.64$, of monodisperse spheres given by
$\varrho^*(4\pi{\left<R\right>}^3 /3) =f$ results in $\phi^* = (3\left<h\right>/4\left<R\right>)f \approx 8.3\%$. This value is
somewhat fortuitously close to the concentration, $\phi_{\textrm{I/LC}} = 8\%$, where the I/LC transition is observed experimentally.

Since the dynamics of freely rotating charged platelets at concentrations well below the overlap concentration $\phi^*$ is most strongly influenced
by the monopolar terms in the far-field electrostatic and hydrodynamic interactions, and since collective diffusion is to a large
extent triggered by the osmotic compressibility,
$D_c(\phi)$ can be expected to behave similar to the collective diffusion coefficient
measured in low-salinity suspensions of spherical charged colloids \cite{Patkowski2005, heinen2010short}.
The latter coefficient also shows a distinct rise at lower $\phi$.

We argue that \figureref~\ref{fig:Modes_vs_phi} depicts the long-time values of the translational self-diffusion coefficient $D_t$
which, for interacting systems, can be substantially smaller than its short-time counterpart, whose deviations from the infinite dilution limit $D_t^0$
are comparatively small. That $D_t$ should be interpreted as a long-time property indeed follows from \expressionname~\eqref{eq:double_exponential}
and the considered small-$q$ range in
combination with the observation that $D_c > D_t^0 > D_t$ at $\phi \gtrsim 0.5\%$. Due to the fast decay of the collective mode, $\Gamma_t^{\text{VU}}$
is determined essentially by the coefficient $D_t$ for times $t \gtrsim 1/(q^2 D_c)$, which is part of the long-time
regime $t \gg \tau_I^0 \approx 0.05~\text{ms}$ once the zero-$q$ extrapolation is made in determining $D_t$.
Different from self-diffusion, the long-time value of $D_c$ is practically equal to the short-time value,
for all concentrations in the isotropic liquid state regime \cite{Nagele1996, Szymczak2004}.

\subsection{Simplifying diffusion model of platelets}

To analyze the trends in the concentration dependence of the gibbsite platelet diffusion coefficients, and to gain further
support for the interpretation of $D_c$ as a collective diffusion coefficient, we introduce here a simplifying model. In this model,
the platelets are described regarding their interactions as uniformly charged effective hard spheres of radius $a$,
interacting by the repulsive electrostatic part, 
\begin{equation}\label{eq:potential} 
\beta u_{\text{el}}(r> 2a) =  L_B Z^2 {\left( \frac{e^{\kappa a}}{1 + \kappa a} \right) }^2 \frac{e^{-\kappa r}}{r}
\end{equation}
of the DLVO potential \cite{Verwey_Overbeek1948}. Here $L_B = 1.18$ nm is the Bjerrum length of DMSO at $T = 293 K$, $Z$ is the effective
particle charge in units of the proton elementary charge, and $\kappa^2 = 4\pi L_B [n_{\text{eff}}|Z| + 2 n_s]/(1-\phi_{\text{eff}})$ is the 
square of the Debye screening parameter. The latter is determined by the number concentration, $n_s$, of residual 1-1 electrolyte
ions which we assume to be equal to $7 \mu$M, and the number concentration, $n_{\text{eff}}|Z|$, of surface-released monovalent counterions.
Furthermore, $n_{\text{eff}} =  3\phi_{\text{eff}}/(4\pi a^3)$ is the number concentration of effective spheres. 
There are several alternatives to define the effective sphere radius $a$, depending on the quantity considered. In the present model,
$a$ is obtained from equating the $2^{\text{nd}}$ virial coefficient of neutral spheres of radius $a$ to that of cylindrical platelets
of radius $\left<R\right>$ and height $\left<h\right>$, assuming all orientations to be equally probable \cite{Isihara1950}.
This results in $a = 0.735\left<R\right> = 32.5~\text{nm}$. A slightly smaller value of
$a = {(3\pi/32)}^{1/3}\left<R\right> = 29.4~\text{nm}$ would be obtained from the $2^{\text{nd}}$ virial coefficient in the zero-$h$ limit.

\begin{figure}
\centering
\hspace{-.5em}\includegraphics[width=.35\textwidth,angle=-90]{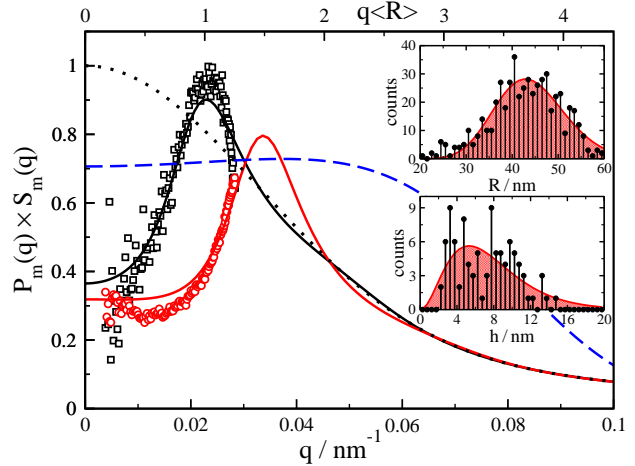}
\vspace{-1em}
\caption{Normalized mean scattered intensity, $I(q)/(I_0 \phi)$, of gibbsite platelets in DMSO, for $\phi = 0.16\%$ (black squares) and $\phi = 0.88\%$ (red circles),
in comparison to the theoretical fits using $\phi_{\text{eff}} = 0.43\%$ (black line) and $\phi_{\text{eff}} = 1.68\%$ (red line).
Black dotted curve: measurable form factor $P_m(q)$.
Blue dashed curve: decoupling amplitude $X(q)$.  
Insets: experimental histograms of the platelet
radius $R$ (top) and thickness $h$ (bottom). The red shaded areas are least-squares fits by unimodal Schulz
distributions with mean values $\left<R\right> = 44.2$ nm and $\left<h\right> = 7.66$ nm,
and relative standard deviations of $s_R = 17.3\%$ and $s_h = 55.3\%$, respectively.
}
\label{fig:Iq_fits}
\end{figure}

In addition to the DLS measurements described before, we have measured the static mean scattered intensity, $I(q)$, using 
SLS. \figureref~\ref{fig:Iq_fits} includes the experimentally determined intensities (open symbols),
for platelet volume fractions $\phi = 0.16\%$ (black) and $0.88\%$ (red).

Light scattering measurements are restricted to a smaller $q$-range than small angle X-ray scattering (SAXS)
so that the interaction peak in $I(q)$ is resolved experimentally
for the lower concentrated system only. Using the effective sphere interaction model, with $u_{\text{el}}(r)$ according to \expressionname~\eqref{eq:potential},
we have calculated $I(q)$ approximately within the rotational-translational decoupling
approximation \cite{Nagele1996,Chen1983}, where it is given by 
\begin{equation}
I(q) =  I_0 \phi P_m(q) S_m(q)\label{eq:Iq_decouple},
\end{equation}
with the so-called measurable static structure factor,
\begin{equation}\label{eq:Sm} 
S_m(q) = \left[1 - X(q)\right] + X(q)S(q).
\end{equation}
Here, $I_0$ is a $q$-independent factor with the dimension of intensity, which is the same in all
intensity measurements corrected for source intensity, recording time,
and scattering volume. Moreover, $S(q)$ is the structure factor of the monodisperse system of effective spheres interacting
by the pair potential in \expressionname~\eqref{eq:potential}. The measurable form factor, $P_m(q)$,
and the dimensionless decoupling amplitude $X(q)$,
\begin{eqnarray}
P_m(q) &=& \left<f^2\right>(q), \label{eq:formfact}\\
X(q)   &=& {\left<f\right>}^2(q)/P_m(q), \label{eq:decoupling_func_X} 
\end{eqnarray}
with $0 \leq X(q) \leq 1$,   
are given in terms of the first and second moments of the scattering amplitude, $f(q,R,h,\mu) = V_p b(q,R,h,\mu)$, of a cylindrical
platelet of volume $V_p = \pi R^2 h$, where $\mu$ denotes the cosine of the angle between a platelet's rotational symmetry axis and
the scattering wave vector. The simple analytic expression for the dimensionless form amplitude, $b(q,R,h,\mu)$, is given in the literature
\cite{Pedersen1997,Kubota1985,Deutch1981}.

\expressionsname~\eqref{eq:Iq_decouple}-\eqref{eq:decoupling_func_X} are based on the assumption that the particle shape, orientation,
and the size polydispersity affect the form amplitudes only, which excludes anisotropic interaction effects. Therefore, the
decoupling approximation can be reasonably applied at lower concentrations only, where the platelets rotate essentially freely.
The brackets, $<\ldots>$, in \expressionsname~\eqref{eq:formfact} and \eqref{eq:decoupling_func_X} represent averages with respect to the platelet
orientation variable $\mu$ (assuming free rotation), and the platelet radius and height joint distribution function $p(R,h)$, which we
assume to be factorisable as $p(R,h) = p_R(R) \times p_h(h)$. Histograms for the marginal distributions $p_R(R)$ and $p_h(h)$ of the dried specimen,
obtained from electron microscopy and atomic force microscopy, respectively, are shown in the insets of \figureref~\ref{fig:Iq_fits}.   
On fitting unimodal Schulz distributions both to $p_R(R)$ and $p_h(h)$, we obtain mean values $\left<R\right> = 44.2$ nm and
$\left<h\right> = 7.66$ nm, at relative standard deviations of $55.3\%$ and $17.3\%$, respectively.
Note that for $q \lesssim 2/\left<R\right>$, $X(q) \approx 0.72$ stays practically constant.

The static structure factor $S(q)$ in \expressionname~\eqref{eq:Sm}, describing pair correlations of effective charged spheres, has been calculated using the
modified penetrating-background corrected rescaled mean spherical approximation (MPB-RMSA), introduced recently by part
of the present authors \cite{Heinen2011,Heinen_Erratum_2011}.
This analytic method allows for an efficient and accurate calculation of structure factors of non-overlapping spheres,
interacting by a repulsive Yukawa-type potential such as the one given in \expressionname~\eqref{eq:potential}.

A fit of the decoupling approximation expression for $I(q)$ in \expressionname~\eqref{eq:Iq_decouple} to the experimental intensities
involves $Z$, $\phi_{\text{eff}}$, $I_0$ and, within a reasonably small range, also $n_s$ as the fitting parameters. Since an
accurate determination of the effective particle charge number $Z$ relies on fitting the maximum in $I(q)$, not located inside
the SLS-resolved $q$-range for most of the considered
concentrations, we alternatively estimate $Z$ using the experimentally observed I/LC transition concentration, determined in our
low-salinity experiments as $\phi_{\text{I/LC}} = 8\%$. Earlier simulations \cite{Frenkel1982,Frenkel1992} predict 
the isotropic-nematic transition of neutral hard disks to occur at 
$n_{\text{I/LC}} d^3 \approx 4$, where $n_{\text{I/LC}}$ is the isotropic-phase number density at the transition point,
and $d$ is the disk diameter.
Using this relation, we estimate the charge on a gibbsite platelet by assuming that $d = 2[\left<R\right> + \kappa^{-1}]$
and $n_{\text{I/LC}} = \phi_{\text{I/LC}}/(\pi \left<R^2 h\right>)$, with $\kappa^2 = 4\pi L_B (n_{\text{I/LC}}|Z| + 2n_s)/(1-\phi_{\text{I/LC}})$.
For the considered large concentration at the I/LC transition point, the residual salt contribution to $\kappa$ can be neglected.
Solving for the effective platelet charge gives the value $Z=71$.
For simplicity, and since the concentration dependence of $Z$ for gibbsite in DMSO is unknown, this effective charge
value is used in all our calculations, independent of $\phi$. 
According to \cite{Franks2006}, $Z = 71$ is a reasonable charge value for gibbsite platelets.

As demonstrated in \figureref~\ref{fig:Iq_fits}, decently good fits of the experimental $I(q)$ in the probed $q$-range are obtained from adjusting the only remaining
fit parameter, $\phi_{\text{eff}}$, entering into the MPB-RMSA calculation of $S(q)$. The global factor $I_0$ in \expressionname~\eqref{eq:Iq_decouple}
only sets the overall intensity scale and is therefore system independent. Assuming a homogeneous linear relation between $\phi_{\text{eff}}$ and $\phi$,
so that $\phi_{\text{eff}}=0$ for $\phi=0$, from our fit we obtain $\phi_{\text{eff}} = 1.93\phi$.

While the theoretical fit of the SLS $I(q)$ is satisfying, future SAXS measurements are required to obtain $I(q)$ in a more extended $q$-range which
covers all its structural features, allowing for a more detailed fitting.

The effective sphere interaction model in combination with the rotational-translational decoupling approximation is easily generalized
from the SLS intensity to the
normalized time-dependent electric field autocorrelation function, which for small $q$ in VV ($\approx$VU) geometry is given by
\begin{equation}\label{eq:EFAC}
g_1^{\text{VV}}(q,t) = [1-B] e^{\displaystyle{-q^2 W(t)}} + B S(q\to0) e^{\displaystyle{-q^2 D_c t}}, 
\end{equation}
with $B = X(0) / [ X(0) + (1-X(0))S(0)]$. Here, $W(t)=1/6 <{[\mathbf{r}(t)-\mathbf{r}(0)]}^2>$ is the mean-squared displacement of a charged
effective sphere at center-of-mass position $\mathbf{r}(t)$ with initial (short-time) slope $D_t^S$,
and final (long-time) slope $D_t^L$, where $D_t^L < D_t^S \leq D_t^0$. Moreover, on
ignoring the very small difference between long- and short-time collective diffusion coefficients, which becomes noticeable at very high concentrations only,
$D_c$ is given by $D_c = D_t^0 H(q\to0)/S(q\to0)$, where $H(q)$ is the so-called hydrodynamic function \cite{Jones1991, Nagele1996, Dhont1996}.
\expressionname~\eqref{eq:EFAC} is fully consistent with the for the gibbsite in DMSO suspensions observed
two-mode decay of the IACFs, fitted using \expressionname~\eqref{eq:double_exponential},
and with the interpretation of the experimental $D_t$ as a long-time self-diffusion coefficient.
We note that in the simplifying model considered here, the diffusion coefficients of platelets are simply approximated by those of the effective charged spheres.

Using the effective sphere parameters $n_s = 7 \mu$M, $a = 32.5$ nm, and $Z = 71$ determined as described above,
and on varying the (effective) volume fraction in
small steps using the relation $\phi_{\text{eff}} = 1.93\phi$, a set of $S(q)$'s has been generated using the MPB-RMSA scheme,
which in turn was employed as input
to the otherwise parameter-free self-part corrected $\delta\gamma$ method of calculating $H(q)$
\cite{Genz1991,Banchio2008,Heinen_TheoArticlePreparation,Beenakker1983,Beenakker1984}.
The corrected $\delta\gamma$ method includes many-body hydrodynamic interactions in an approximate way, making predictions for the $H(q)$ of
Brownian spheres with Yukawa-type repulsion in overall good agreement with Stokesian Dynamics computer simulation results \cite{Heinen_TheoArticlePreparation}.

Regarding the approximations invoked in our simplifying analytic model, the agreement between the calculated $D_c(\phi)$ and
the experimental data is rather satisfying.
The increase of $D_c$ with increasing $\phi$ to values about $20$ times larger than $D_t^0$ is qualitatively captured. The theoretical $D_c$ reaches a
shallow maximum at about $\phi = 3\%$, originating from the interplay of osmotic compressibility and sedimentation coefficient
$H(q\to0)$ as discussed, \textit{e.g.}, in \cite{Patkowski2005,Heinen_BSAArticlePreparation}. 

The large-$q$ limiting value of $H(q)$ is equal to the reduced short-time self-diffusion coefficient $D_t^S/D_t^0$. The self-part corrected
$\delta\gamma$-scheme result for $D_t^S(\phi)$ is included into \figureref~\ref{fig:Modes_vs_phi} (see also the inset).
In the explored $\phi$-range, $D_t^S(\phi)$ decreases only mildly with increasing $\phi$. Its $\phi$-dependence is overall well described by the form 
$(1 - a_t\phi_{\textrm{eff}}^{4/3})D_t^0$ with $a_t \approx 2.9$, which is the typical concentration-dependence of $D_t^S$
for low-salinity systems of charged spheres \cite{heinen2010short}. For $\phi \gtrsim 2\%$, however, the decay of the experimental (long-time) $D_t$ in
\figureref~\ref{fig:Modes_vs_phi} with increasing $\phi$ is very strong. For volume fractions $\phi \gtrsim 7\%$ near to $\phi_{\text{I/LC}}$,
$D_t$ has decreased to values less than $1\%$ of $D_t^0$.
This is similar to the slowing down of self-diffusion seen for
block copolymers in a neutral solvent when the disordered-ordered transition is approached \cite{Holmqvist2003}, 
in polymer solutions with increasing concentration \cite{Ballauf2009}, and also for polymer grafted clay particles \cite{Giannelis2005}.
The measured values of $D_t^L/D_t^0$ for gibbsite are much smaller than those reached by the long-time translational
diffusion coefficient of low-salinity charge-stabilized spheres in the fluid phase, which reaches its minimal value of $D_t^L/D_t^0 \approx D_t^L/D_t^S \approx 0.1$ 
at the freezing transition volume fraction $\phi_f$ \cite{Loewen1993,Bergenholtz2000}. For the effective sphere model used here, $\phi_f$ can be estimated 
on basis of the empirical Hansen-Verlet freezing criterion $S(q_m, \phi_{\textrm{eff}} = \phi_f) \approx 3.1$
for charged spheres at low salinity \cite{StevensRobbins1993,HansenVerlet1969,KremerGrest1986}.
With $S(q)$ computed in MPB-RMSA, this results in the freezing transition volume fraction $\phi_{\textrm{eff}} = 42\%$,
corresponding to $\phi = \phi_{\textrm{eff}}/1.93 = 22\%$. 

We attribute the strong decay of the experimental $D_t$ at larger $\phi$ to the
uprising influence of the anisotropic electro-steric interaction parts, and to the hydrodynamic rotational-translational coupling effects in platelet
systems. These anisotropic interaction effects, not included in the effective sphere model, cause an additional strong slowing down of the
translational and rotational self-dynamics.
The inset in \figureref~\ref{fig:Modes_vs_phi} displays the experimental data for $D_t/D_t^0$ (and $D_r^\perp/D_r^{0,\perp}$) on a smaller
concentration range $\phi \leq 0.025$, in comparison with the corrected $\delta\gamma$-scheme predictions for $D_t^S/D_t^0$,
and simplified mode-coupling theory (MCT) results for $D_t^L/D_t^0$ within the effective charged sphere model. In the simplified MCT result,
the enhancing influence of HIs on $D_t^L$, typical of low-salinity systems, is accounted for \cite{BaurEPL1997, Baur1997, Maret1997}. 
Note here that the peak height, $S(q_m) = 1.7$ of the static structure factor at $\phi_{\textrm{I/LC}} = 8\%$ is still rather close to one so that our
usage of the simplified MCT solution in place of the fully self-consistent MCT solution for $D_t^L$ is justified.

The short-time rotational self-diffusion coefficient, $D_r^S$, in charged-sphere colloidal systems follows at low salinity
to a good accuracy the scaling relation $D_r^S/D_r^0 = 1-a_r\phi_{\textrm{eff}}^2$, with $a_r \approx 1.3$, in the whole $\phi$-range covered in
\figureref~\ref{fig:Modes_vs_phi} \cite{Banchio2008, Koenderink2003}. This curve is shown as the solid blue line 
in the inset of \figureref~\ref{fig:Modes_vs_phi}. Different from platelets with $h>0$, spheres
are characterized by a single zero-concentration
rotational diffusion coefficient $D_r^0 = k_B T /(8\pi \eta_0 a^3)$. Note here that $D_r^S/D_r^0$ decreases less strongly with 
increasing $\phi$ than the experimental $D_t^S/D_t^0$. This difference originates from the shorter-ranged hydrodynamic self-coupling of rotational motion \cite{Koenderink2003}.
The experimental diffusion coefficients $D_t$ and $D_r^\perp$ depicted in \figureref~\ref{fig:Modes_vs_phi}, obey overall the
same ordering relation $D_r^\perp(\phi)>D_t(\phi)$ as their short-time counterparts in the effective sphere model.

The low-$q$ expression for $g_1^{\text{VH}}(q,t)$ in our simplifying model is given by (see, e.g., \cite{Zhang2002,Jones1995})
\begin{equation}\label{eq:VH_EFAC}
g_1^{\text{VH}}(q,t) = e^{-q^2 W(t)} G_r(t), 
\end{equation} 
where $G_r(t) = < P_2(\hat{\mathbf{u}}(t)\cdot\hat{\mathbf{u}}(0)) >$ is the rotational self-dynamic correlation function
of spheres, with the optical axis of a sphere characterized by the unit vector $\hat{\mathbf{u}}$, and with $P_2$ denoting the $2^{\text{nd}}$-order
Legendre polynomial. In the derivation of \expressionname~\eqref{eq:VH_EFAC}, it has been assumed that the translational-rotational motions
of a particle are decoupled \cite{Jones1995}. This decoupling is exactly valid for hydrodynamically interacting spheres at short times only.
For non-spherical particles, it is an approximation even to linear order in $t$.

At short times, $G_r(t) = \exp\{-6 D_r^S t \}$ decays exponentially. At long times, however, $G_r(t)$ decays in principle 
non-exponentially, with an average decay rate somewhat smaller than $D_r^S$ \cite{Jones1995}.
While a genuine long-time rotational self-diffusion coefficient does not exist, one can define instead a mean 
orientational self-diffusion coefficient, $\overline{D_r}$, determined by the time dependence of $G_r(t)$ for all times.
A corresponding $\overline{D_r}$ of platelets is shown indeed in \figureref~\ref{fig:Modes_vs_phi}, as obtained
in panel (d) of \figureref~\ref{fig:Autocorr_functions} using the KWW analysis. The calculation of $\overline{D_r}$ for colloidal hard spheres
in \cite{Jones1995} suggests that $\overline{D_r}$ is only slightly smaller than $D_r^S$, at least for smaller values of $\phi$. The 
mean rotational diffusion coefficient $D_r^\perp$ depicted in \figureref~\ref{fig:Modes_vs_phi}, however,
decreases very strongly at larger $\phi$, to an extent comparable to that
of $D_t$. Like for $D_t$, we attribute this strong decline of $D_r^\perp$ at larger $\phi$ to the strong anisotropic 
electro-hydrodynamic coupling of the charged platelets.    

\subsection{Static viscosity results}

\begin{figure}
\centering
\includegraphics[width=.39\textwidth,angle=-90]{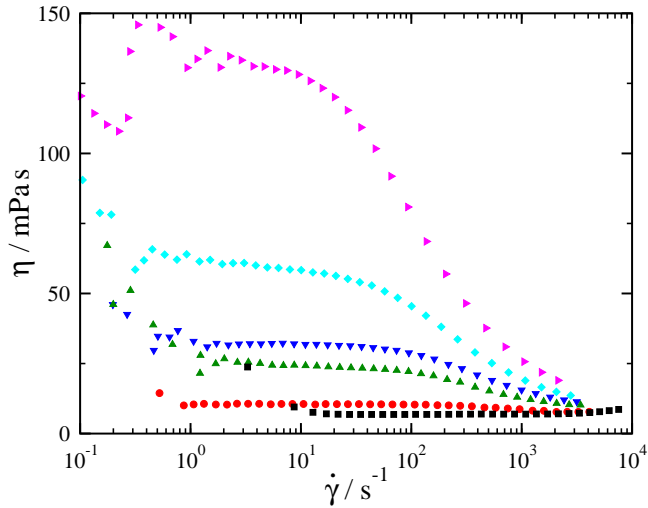}
\vspace{-1em}
\caption{Shear viscosity $\eta$ of gibbsite platelets in DMSO versus shear rate $\dot{\gamma}$, for $\phi = 7.6\%$, $6.9\%$, $5.4\%$,
$4.0\%$, $2.4\%$, and $1.45\%$ (from top to bottom).}
\label{fig:visco}
\end{figure}

It is interesting to note from \figureref~\ref{fig:Modes_vs_phi} that the strong decline of the experimental $D_t/D_t^0$ and $D_r/D_r^0$
with increasing $\phi$, is qualitatively similar to the concentration dependence
of the normalized inverse low shear-rate limiting viscosity, $\eta_0/\eta$.
To obtain the low-shear rate viscosity data shown in \figureref~\ref{fig:Modes_vs_phi},
using an ArG2 rheometer from Ares, we have measured the viscosity as a function of shear rate
$\dot{\gamma}$ for various concentrations.
Results of these measurements are depicted in \figureref~\ref{fig:visco}, not unexpectedly showing that
the gibbsite suspensions are shear-thinning, with the shear-thinning regime shifting to lower shear rates with increasing
volume fraction. The low shear-rate limiting viscosity $\eta$ is identified by the (mean) Newtonian plateau value
at low shear rates, with results depicted by stars in \figureref~\ref{fig:Modes_vs_phi}. For technical reasons,
no reliable viscosity data have been obtained for very low shear rates $\dot{\gamma} \lesssim 1 s^{-1}$.
Moreover, for technical reasons our viscosity data do not include the low-concentration regime from which the $\phi$-independent intrinsic viscosity
,$[\eta] = \lim_{\phi\to0} (\eta- \eta_0)/(\eta_0 \phi)$, could be determined, which depends on $p$ only.
For a sphere $(p = 1)$ with stick hydrodynamic boundary conditions, $[\eta] = 2.5$, and larger values for $[\eta]$ are obtained with
increasing asphericity. From the power-law
representation of the simulation data in \cite{DeLaTorre2003} for the intrinsic viscosity of thin cylinders, $[\eta] \approx 6.8$
is obtained for $p = 0.087$. This value is similar to the one observed in \cite{Philipse2001}.
Note that a spheroid of equal aspect ratio has a distinctly larger intrinsic viscosity of $[\eta] \approx 9.1$.


\section{Conclusions}\label{sec:conclusions}

Using (D)DLS and SLS, we have measured the long-time translational collective- and self-diffusion
coefficients $D_c$ and $D_t^L$, the mean rotational diffusion coefficient $D_r^\perp$, and the static scattered intensity $I(q)$
of charged gibbsite platelets in DMSO at low ionic strength. Our experiments cover the concentration range from very dilute systems up to the I/LC transition.

Our usage of DMSO as a solvent with a dielectric constant close to that of gibbsite,
has enabled us to determine the translational and rotational diffusion properties without the necessity
of invoking elaborate X-ray photon correlation spectroscopy measurements. 
A fast relaxation mode in the dynamic scattering data has been frequently reported in relation
to the liquid-glass transition, the glassy state and also for polymer coated clays \cite{ Nicolai2001, Abou2001, Bonn2004}.
In the present study, a fast mode has been found also for the isotropic phase, and it has been identified as a collective diffusion mode.

With increasing $\phi$, the measured collective diffusion coefficient increases up to about twenty times
the single-particle (orientationally  averaged) translational diffusion coefficient.
Different from the translational and rotational self-diffusion coefficients, which strongly decrease for
$\phi \gtrsim 2\%$, $D_c$ stays nearly constant for these larger concentrations even up to the I/LC transition concentration.
The strong decay of the self-diffusion coefficients is accompanied by a comparatively pronounced increase
of the zero-shear-rate limiting static shear viscosity. We have provided arguments, both experimentally and theoretically,
that the coefficients $D_t$ and $D_r^\perp$ obtained in our scattering modes analysis, should be identified,
respectively, with the translational long-time and the mean rotational self-diffusion coefficients of gibbsite platelets.

At low concentrations, $I(q)$ is well reproduced in the simplifying translational-rotational decoupling method, 
where correlations between particle positions, sizes and orientations are neglected, and where the direct platelet interactions are approximated
by a spherically symmetric electrosteric repulsion of DLVO type. The effect of hydrodynamic interactions is accounted for in our effective sphere model.

Except for very low concentrations, the accessible $q$-range in light scattering experiments is restricted to 
wavenumbers smaller than the value where the principal structure peak in $I(q)$ occurs.
The effective particle charge has therefore been estimated, from the I/LC transition concentration, as $Z = 71$, and was kept constant in our
model calculations independent of concentration and salinity.
For an unambiguous determination of $Z$, and to explore its dependence on $\phi$ and $n_s$, a broader $q$-range than accessible in SLS
is required, which can be studied in future SAXS and XPCS measurements. In a more refined theoretical model for $I(q)$, one can 
account for shape-dependent direct interaction contributions using the PRISM model \cite{Andersen1970, Hansen_McDonald1986, Schweizer1997, Harnau2002}.
However, regarding the dynamic quantities, it will be very difficult to include shape-dependent hydrodynamic interactions,
on avoiding numerically expensive multiparticle collision, fluctuating Lattice-Boltzmann or Stokesian dynamics simulations.
We note here that while the simple effective sphere model clearly fails in terms of quantitative predictions,
it allows for a correct assessment of qualitative features in the isotropic phase such as the ordering in magnitude of the rotational
and translational self-diffusion coefficients, and the approximate plateau region of the collective diffusion coefficient at larger $\phi$.

Upon increasing the concentration, peculiar observations have been made in the measured diffusion properties of the gibbsite/DMSO system. 
Similar to suspensions of charge-stabilized colloidal spheres at low ionic strength,
$D_c$ grows with increasing concentration. Since the value of $D_c$ is basically the same
at long and short time scales, and since the increase in $D_c(\phi)$ is mainly due to the reducing
osmotic compressibility $S(q\to0)$, these basic features of $D_c(\phi)$ have been reproduced
in our effective sphere model, using the calculated $S(q)$ as input to the otherwise parameter-free self-part corrected $\delta\gamma$-scheme
describing the short-time dynamics.

The employed effective sphere model of gibbsite is less accurate regarding the measured
long-time dynamic quantities $D_t^L$ and $D_r^\perp$, which are more sensitive to the anisotropic direct and hydrodynamic interaction
parts most influential on shorter length-scales. The strong decay of both self-diffusion coefficients
to less than $1\%$ of their respective infinite dilution values near 
$\phi_{\textrm{I/LC}} = 8\%$ is not reproducible in an effective sphere model, where $D_t^L \gtrsim 0.1 \times D_t^0$
even at the largest fluid-state concentration.  
      
A concentration-dependence similar to the ones of $D_t^L$ and $D_r^\perp$ has been found for the inverse low shear-rate limiting viscosity
${(\eta/\eta_0)}^{-1}$. Examinations of possible generalized Stokes-Einstein relations between viscosity and the various diffusion coefficients
in concentrated platelet fluids could be the topic of a future study based on the present work. 


\section*{Acknowledgments}
P. Davidson and H.H. Wensink are thanked for enlightening discussions. The work of D.K. was financed by the
Foundation for Fundamental Research on Matter (FOM), which is part of the Netherlands Organization for Scientific Research (NWO).
M.H. acknowledges support by the International Helmholtz Research School of
Biophysics and Soft Matter (IHRS BioSoft). 
G.N. acknowledges funding from the Deutsche Forschungsgemeinschaft (SFB-TR6, project B2).

\end{document}